**Title:**

Testing Human Ability To Detect "Deepfake" Images of Human Faces

Contributors:

Sergi D. Bray[1,*], Prof. Shane D. Johnson[1], Dr. Bennett Kleinberg[2]

[1]Department of Security and Crime Science, University College London (UCL), WC1H 9EZ, UK.

[2]Department of Methodology, University of Tilburg, 5037 AB Tilburg, The Netherlands.

* - Corresponding address: 35 Tavistock Square, London, WC1H 9EZ, UK.
Email: sergi.bray.18@ucl.ac.uk Telephone: +44 (0)20 3108 3206

**Abstract**

"Deepfakes" are computationally-created entities that falsely represent reality. They can take image, video, and audio modalities, and pose a threat to many areas of systems and societies, comprising a topic of interest to various aspects of cybersecurity and cybersafety. In 2020 a workshop consulting AI experts from academia, policing, government, the private sector, and state security agencies ranked deepfakes as the most serious AI threat. These experts noted that since fake material can propagate through many uncontrolled routes, changes in citizen behaviour may be the only effective defence.

This study aims to assess human ability to identify image deepfakes of human faces (these being uncurated output from the StyleGAN2 algorithm as trained on the FFHQ dataset) from a pool of non-deepfake images (these being random selection of images from the FFHQ dataset), and to assess the effectiveness of some simple interventions intended to improve detection accuracy. Using an online survey, participants ($N$=280) were randomly allocated to one of four groups: a control group, and three assistance interventions. Each participant was shown a sequence of twenty images randomly selected from a pool of 50 deepfake images of human faces and 50 images of real human faces. Participants were asked whether each image was AI-generated or not, to report their confidence, and to describe the reasoning behind each response.

Overall detection accuracy was only just above chance and none of the interventions significantly improved this. Of equal concern was the fact that participants' confidence in their answers was high and unrelated to accuracy. Assessing the results on a per-image basis reveals that participants consistently found certain images easy to label correctly and certain images difficult, but reported similarly high confidence regardless of the image. Thus, although participant accuracy was 62% overall, this accuracy across images ranged quite evenly between 85% and 30%, with an accuracy of below 50% for one in every five images. We interpret the findings as suggesting that there is a need for an urgent call to action to address this threat.

**Data and code availability**
The data and code of this study are available at: https://osf.io/tfn7v/

## 1. Introduction

A "deepfake" is an entity that is created by complex algorithmic computation with minimal, if any, human supervision (hence "deep") that falsely represents reality ("fake") [1–3][1]. "Shallowfakes" or "cheapfakes" are, in contrast, entities falsely representing reality that have been created by a human [11,12]. Deepfakes are considered to fall under the umbrella term of "synthetic media", although given the field's nascent state and continuing development there has not been a clear positioning of terms in recent literature. A deepfake may take many forms and is not confined to being a digital entity: an image generated by an artificial intelligence (AI) may be printed on paper, for example. Prior literature has suggested that deepfakes fall into three major categories: head puppetry, face swapping, and lip syncing [13], but the situation is more complex than this summary portrays. Rather than give a fuller taxonomy, we attempt in this introduction to explain the relevant elements at a conceptual level.

Deepfake generation processes can be conceptually identical to processes that were previously possible, such as the seamless transplantation of one human face onto another within video footage. This example is achievable either through a manual or computational process that involves cutting and pasting, frame-by-frame, parts of existing footage onto the target footage [14]. While possible, it would be impractically time-consuming to do this at scale manually but this process can be automated using AI. However, deepfakes can also take forms that could not result from manual processes.

One example of a deepfake is the "style transfer" type of deepfake [15–19]. The deepfake generation method for this type of deepfake uses machine learning to conduct pattern-recognition [20] across, for example, a very large dataset of images of human faces. The patterns it is programmed to recognise, essentially, are the patterns that make the image of a human face exactly that: an image of a human face. As such, the Machine Learning system can then, using these patterns that it has learned to recognise, create its own images of human faces. To be clear, these images are not images of any of the human faces that were given to the system as input; the system has learned what human faces look like, and has started generating its own images that effectively tick the boxes for "looking like" an image of a human face. This example is about images of human faces, but any dataset can be given to this type of deepfake generator, and new deepfake instances can be generated in the "style" of any input dataset. The word "style" here is tricky, and should be understood as a human way of describing in abstraction a complex algorithmic process. Portrait images may represent one "style", while profile images will represent another. This deepfake generation method, of creating a new, deepfake instance in the "style" of a given dataset – quickly and at scale – does not have an achievable analogue equivalent.

---

[1] This is the view of the authors of this paper. The words "entity" and "reality" have been chosen as they are broad enough concepts to encompass the current and future purview of what a deepfake can be. We acknowledge that this is a novel definition and as such may be contested especially where one is more minded towards descriptivism than prescriptivism [4,5]. A descriptivist might look to definitions given by other people, for instance the anonymous author(s) of the Wikipedia article on deepfakes [6] who limit the term's envelope to videos or images in which one person's likeness has been replaced by another person's. A prescriptivist approach would step back from common usage to consider how the term should be used, from an a priori standpoint. This is the approach we take, and we argue that there are benefits to this approach in that a conceptual definition is flexible enough to allow future development in this research area to fall under the "deepfake" definitional umbrella. The descriptivist approach would mean that definitions must be continually updated, and can lead to problems when updates are not correctly merged: the existence of a detached Wikipedia entry on "Audio Deepfake" [7] is confusing if not confused when the main "Deepfake" entry [6] limits its definition to videos and images. Whether DALLE-2 or ChatGPT-3 are deepfake generation technologies is, for the descriptivist, currently a question of which sources to listen to [8,9]. The prescriptivist approach we take in giving our definition incidentally mirrors recent research [10] where, although no definition is given, the meaning of "deepfake" encompasses multiple modalities and indeed breaks from the Wikipedia definition by acknowledging "puppet master" deepfakes, where a person's likeness is puppeteered rather than replaced by a different person's likeness. We hope that this footnote assuages any criticism of our definition but understand if readers are differently minded than ourselves.

The central basis behind our preference for a conceptual approach to the definition of deepfakes is that the core technology that deepfakes are generated by is machine learning, also known as deep learning [21] (this is where the first half of the portmanteau "deepfake" comes from). There are several strands of machine learning technology, some of which are used to generate deepfakes (Generative Adversarial Networks [22], Auto-Encoders [23], Diffusion Models [24]). Machine learning takes data as input, and the nature of its output depends fundamentally on the nature of this input data. As such, deepfakes generated by these methods will derive much of their attributes from the input data that the machine learning process has received. There may be empirical nuances that prohibit certain instances, but on an abstract and conceptual level, machine learning technology is agnostic with respect to the modality of the input data and to the (human-perceptual) content of that input data. The import of this is that the content of a deepfake could theoretically be anything within a given modality (video, image, audio, or text) [25–32]. The existing literature shows that deepfakes can, at least in one modality, portray humans [33,34], art [18], animals [35], landscapes [18], food [18], satellite imagery [18], street maps [18], room interiors [36], and dashcam footage [36]. Deepfakes have positive applications. For example, audio deepfakes of people's voices can create artificial voice replacements for patients whose ability to speak is taken from them by motor neurone disease [37]; inter-language dubbing in films could be accompanied by deepfake versions of the actors in which their mouths are animated to match the new, dubbed, words [38]. It could also be claimed that deepfake technology has artistic applications [39,40]. Such claims have been questioned by experts including senior art critics, on the grounds that the created content can be "bland" or "boring", and lacks "originality" [41,42]. Beauty is in the eye of the beholder, of course, and there also appear to be hybrid artistic applications. For example, composers have begun to splice samples of AI-generated music into their compositions, combining their human imagination with AI-generated inspiration to produce a novel, cyborg art form [43]. However, while positive use cases exist, there clearly are harmful applications, such as the manipulation of specific visual segments of surveillance camera footage in real-time [25]. In addition, there appear to be use cases that need further research before implementation to verify whether their purportedly positive aspects are not in fact harmful in the long term [44,45].

Perhaps most disturbing is the fact that deepfakes enable extensions of existing criminal threats ("cyber-enabled" crimes), and new vectors of crime ("cyber-dependent" crimes) [46,47]. Also, while the widescale use of shallow-fakes is limited by the time that it would take a human to produce them, a critical aspect of deepfakes is their scalability: the lack of human involvement in the creation process means that a computer (or a botnet [48]) can be set up to do the intensive computation needed to create immense quantities of deepfakes quickly. In combination with the realism of some deepfakes this creates an as yet unmapped threat potential which can be conceptually fitted to any domain or context.

Like the problems of phishing emails and illegal robocalls we experience today [49,50], this scalability means that a single malicious use of deepfakes could affect large numbers of people, and could move victim targeting from an individual to a societal scale. The prevalence of advertising technologies which allow advertisers to target their product's most relevant audience could also enable attackers to easily locate specific groups of people that they wish to attack online.

A growing body of literature [51–54] including independent reports and whitepapers from the UK Government and Europol [37,55,56] as well as warning broadcasts from the FBI [57], provisions in the US National Defense Authorization Act 2020 [58], and US Army development of deepfake detection algorithms [59], has pointed to the potential harms deepfakes pose. For example, in a recent study on 'AI and Future Crime' [46], 31 representatives with expertise in AI from academia, policing, government, the private sector, and state security agencies, rated deepfake technologies to be the top-ranking threat associated with AI. Crime applications of deepfakes included fraud (such as the "grandparent scam"[2] [60]), authentication forgery to gain access to secure systems, and fake video evidence of public figures speaking or acting reprehensibly in order to manipulate support. The study concluded that, since algorithmic methods of detecting deepfakes may not be possible in the longer term, and because fake material can propagate through many uncontrolled routes, changes in citizen behaviour may be the only effective defence, at least for now.

---

[2] This is an example of a virtual kidnapping which may take various forms. In one scenario, offenders call an individual claiming to have kidnapped their child or grandchild and threaten to harm them should a ransom not be paid. In reality, the child has not been kidnapped and so while the offence is very low cost it can be very profitable to offenders.

A crucial element of the deepfake threat is that it deceives a human actor into believing that it correctly represents reality, rather than merely (and explicitly) mimicking it. At the time of writing, technical solutions are not sufficiently developed to address the deepfake threat, although considerable effort is being invested [61–63]. However, few studies have inspected the extent to which a given deepfake instance actually succeeds in deceiving human perceivers in this way, and there are limitations associated with these studies that are discussed below. Moreover, given that perceptible imperfections exist in some deepfake creation algorithms [64] it is possible that simple human-oriented solutions to the deepfake threat might work. These include the provision of advice on what imperfections to look for. Such interventions might reduce at least some of the risks deepfakes pose, and provide simple scalable solutions. However, whether such interventions would be effective is an empirical question.

The aim of this study was to assess people's ability to differentiate between deepfake (AI-generated) and authentic images of people, and to test whether advice about how to detect a deepfake improves performance. The remainder of this article is organised as follows. In the next section we survey the wider context that the study pertains to, this being the proliferation online of purportedly-true false content. We then outline the literature that has proposed and designed solutions to the deepfake problem, and assess the likely effectiveness of these solutions from a conceptual perspective. Next, we inspect one deepfake generation method in detail, which is the method tested in this paper. We then discuss the Methods employed, present our findings, and discuss the results in the wider context of the research.

### 1.1. Wider Context of Fake Online Content

The negative effects of internet falsehoods can be seen in the increasing distrust people express in news distributed online. For example, according to 2020 statistics from Reuters, 56% of people (sampled across 40 countries) are concerned about the veracity of news found online [65]. Statistics also suggest an erosion of trust in the news more generally, which fell from 44% in 2018 to 38% in 2020 [65,66]. However, these low levels of trust pale by comparison to people's trust in news encountered on social media (23% in 2018, 22% in 2020) and even in news retrieved using search engines (34% in 2018, 32% in 2020) [65,66].

Fake online content has the potential to intensify a plethora of existing problems. From a global perspective, fake content has already contributed to online interference associated with international politics. For example, in 2018, the Oxford Computational Propaganda Project found evidence of organised social media manipulation campaigns in 48 countries [67]. Many such campaigns use disinformation tactics in which fake content may be used to variously discredit, confuse, or create information cascades by outnumbering authentic sources of real content [68], which can alter the direction of history by influencing election results [69]. Advances in fake content generation technologies will likely intensify the use and effect of fake content and possibly increase the (mis)use cases.

Fake online content (especially advances in fake content generation technology) also has the potential to exacerbate a number of wide-reaching societal problems, the effects of which have already been intensified by the internet. For example, the abuse and harassment of public figures, or members of certain demographic groups, including women and minorities, is a critical problem that causes harm at both individual and societal levels by silencing certain voices and can dissuade individuals from public-facing careers [55,70–72], amongst other things.

### 1.2. Proposed Solutions to Deepfakes

A number of solution systems have been proposed to address the threats that deepfake technologies pose. To date, these mainly leverage either the pattern recognition powers of machine learning systems to detect deepfake instances in situ [73], or the ability of Blockchain technologies to create white-lists of non-deepfake instances [74,75]. In the former case, machine learning classifiers use pattern recognition to effect deepfake detection. The method by which they do so is (in simplified form) to give a classifier a labelled training dataset (with labels "real" and "deepfake" applied correspondingly to the stimuli). The classifier is then trained to distinguish between real and deepfake stimuli. These approaches have been shown to work well for test datasets that are similar in content to the training dataset [76–78] but the difficulty will be to find a distinction between the real and deepfake halves of the training dataset that also applies, or in technical terms "generalises to", other datasets (for example with different content, or where a different deepfake generation method has been used to create the deepfake stimuli) [30,79].

In the latter case, the idea has been put forward that the Blockchain can help to distinguish deepfake stimuli from authentic stimuli [75,80–82]. The idea is that either authentic or deepfake instances, or (for instance) hash fingerprints thereof, could be put on the Blockchain public ledger. This would allow people to check if the instance which they suspect of being a deepfake has been referenced on the Blockchain. In this way the Blockchain public ledger could act as, for instance, malware fingerprint storage in antivirus software: a database is formed of real, or of deepfake, instances; the veracity of a novel stimulus is established by checking if this stimulus matches any in the database. There are certain problems with this as a solution that could be relied on in anything other than very limited settings. Presuming a scenario where the threat is an attacker who would create and use novel deepfake instances, this novel deepfake instance would not have yet been added to the Blockchain, and so would not be matched to anything on the Blockchain. This is a central, fundamental and definitive problem. The most that can be given by this method is a storage of "seen-so-far", which could be useful against non-"zero-day" deepfake instances. However, given that the nature of deepfake generation methods is such that they are highly scalable (being fundamentally dependent only on computational resources) there is little motivation for an attacker to re-use a particular deepfake instance when they can churn out thousands of similar deepfake instances in a small amount of time. Moreover this proposal of a Blockchain-based solution reduces to the task of establishing that a given instance is a deepfake or non-deepfake, a task which could either be undertaken by deepfake detection algorithms, or by humans. A similar but distinct option could be to create white-lists of trusted user accounts, which would then be trusted not to upload deepfakes. Problems with this option are that user accounts can be taken over [83], users may be bribed or blackmailed, or untrustworthy users may make it onto the white-list. This option could additionally be undermined by human inability to reliably distinguish a deepfake from a non-deepfake.

Moreover, deepfake detection algorithms will suffer from problems inherent to the task: deepfakes are improving all the time, they can take a variety of modalities, and can comprise a range of content (see above). Algorithmic efforts involve the use of training data to teach Machine Learning systems what a deepfake is. At present, the diversity of the training set used dictates the scope of its effectiveness [84]. For example, a deepfake detection algorithm that succeeds in identifying deepfake images of human faces will not know what features to look for in images of horses – unless it has found features in deepfake images of faces that successfully generalise to deepfake images of horses. The same may apply to other types of human faces if the training and real-world images differ in meaningful ways (e.g. portrait versus profile images). Even if such generalisation is successfully found and capitalised upon, it may be undermined either by human removal of these distinguishing features in image editors (such as Adobe Photoshop [85] or GIMP [86]), or by development of adversarial machine learning technology [87]. There are two key types of adversarial machine learning technology: the first, "evasion attacks", can bypass machine learning systems by sculpting the malicious data instance (in this context, a deepfake) in such a way that it is mis-labelled (i.e. as non-deepfake) by the machine learning system [88]. The second type, "poisoning attacks", works only against "live" (or "unsupervised") machine learning models that update themselves based on the data instances that they observe in the wild [89]. A classic example is the Twitter bot which attackers trained to output increasingly racist remarks [90]. Samples correctly belonging to label A can be made to look increasingly similar to samples of label B, until a systematic mis-labelling of instances is achieved [91]. This is all to say that the development of deepfake detection algorithms is likely to be closely followed by the development of adversarial machine learning technology that reduces or nullifies their efficacy. Such adversarial development may take place out of sight, meaning that threat actors could be achieving their goals silently and successfully without anything looking out of place. Development of these opposing technologies will very likely devolve into a potentially interminable arms race, following precedents in areas such as malware detection, anti-spam, anti-adblocking, and (offline) vehicle crime prevention [92–96].

Deepfake detection algorithms will also struggle to achieve sufficiently protective implementation. The contexts into which such algorithms should be instantiated (for instance to protect social networks from deepfakes) may be managed by authorities, owners, or stakeholders that resist the correct, independent, and transparent implementation of deepfake detection algorithms. Governments and regulatory bodies may be able to enforce this implementation, but implementation should be recognised as complex: deepfakes can take the form of digital but also analogue media – for instance a photo can be printed out onto paper or card. Handheld deepfake scanner technology could be part of a solution, but any solution must recognise that contexts contain complexity and that "Security is a process, not a product" [97].

Of course, the possibility exists to develop support systems that involve humans and technological solutions to detect deepfakes. Researchers [98] have for instance suggested that content moderation systems with human-AI collaboration could be a solution in this area. However, this possibility should not be considered a straightforward one, since even if humans learn from AI predictions as to whether a given instance is a deepfake instance or not, research has not examined if this helps (or hinders) if the predictions of the AI are incorrect. Given the plethora of problems that face deepfake detection algorithms, and the contextual complexity of implementing any algorithm, with or without a human in the loop, within actual defensive security situations, any claims that such solutions are available to implement should be made carefully and investigated thoroughly before use.

One angle rarely explored is testing, and improving, the human detection of deepfakes. Many deepfakes have tell-tale signs that are visible to the human eye, and hence the possibility exists that humans could detect them if they knew what, specifically or generically, to look for. If possible, this could provide a scalable solution that could work in concert with technical measures. On the other hand, if technical measures do not prove to be viable, or until they are developed in a scalable and reliable form, this non-technical measure could present the only viable mitigation to this new threat, at least for now.

### 1.3. Human Detection Ability Studies: Image Deepfakes

Three studies have so far been conducted on human detection ability with respect to static image deepfakes, which are the focus of this paper. In the first of these [61], the study began with a familiarization step where each participant was shown labelled real and deepfake images and given time to inspect them. This was followed by the main experiment where participants were instructed to label a set of new images, one at a time, as "real" or "deepfake". For deepfake images, the study used a mixed dataset of (A) images generated by the StyleGAN1 algorithm [99] which was trained on the Flicker Faces High Quality (FFHQ) dataset [100], (B) images generated by that same algorithm but as trained on the CelebA-HQ dataset [101], and (C) images generated by the PGGAN algorithm [102] as trained on the CelebA-HQ dataset. The real face dataset used images from the FFHQ, Celeb-HQ [103], and CelebA-HQ dataset, although the authors do not state whether these dataset segments (A/B/C) were used in distinct experiments or all pooled into one dataset and used across all experiments. Results suggested that the StyleGAN1 algorithm as trained on the FFHQ dataset produced the most convincing deepfake images of the three techniques: with participant labelling accuracy being 63.9% for (A), 75.15% for (B), and 79.13% for (C).

However, there are some aspects associated with the design of this study which make it difficult to draw conclusions about the reported findings with respect to human performance. First, the authors of the study do not report standard errors or inferential statistics and so it is not possible to determine if accuracy varied in a reliable way across the three deepfake dataset conditions. The small sample size used (N=20) makes this particularly difficult to assess. Second, a great deal of familiarisation is given to participants but it cannot be concluded whether this familiarisation had any effect or not on participants' accuracy because the experimental setup did not include a control group that did not receive this familiarisation. Moreover, the familiarisation given to participants may have been too much to absorb – each participant was shown 10,000 real and 10,000 deepfake images, and was then asked to label 1,000 new images. Viewing such a large number of images would have been cognitively taxing and participant fatigue cannot be ruled out as having affected the results. Furthermore, the mean time that participants took to label each image was only 5.14 seconds. This is quite brief and might suggest that participants were not giving each decision their fullest attention. Finally, the study also does not state how participants were recruited and so it is difficult to know if the findings are generalisable.

In the second such study [104], all participants were provided with a definition of synthetic faces and then completed the experimental task which consisted of sequentially labelling 128 images, pulled randomly from a pool of 800 image stimuli (of which half were real and half deepfake). The first experiment (*n*=315), effectively a control, gave no other training than this definition. In the second experiment (*n*=219), participants were shown a short tutorial describing examples of specific rendering artifacts that could be used to identify fake, synthesized faces from real faces. In this second experiment, participants were additionally given feedback after each response as to whether that response was correct or not. Participants in the second experiment had an average accuracy of 59.0%, while participants in the first experiment attained only 48.2% accuracy, suggesting an improvement in performance for participants in experiment 2. However, participants were not randomly allocated to experiments 1 and 2 which means that they should not be directly compared. To explain, absent random allocation, it is possible that the two groups differed in important ways other than the specific tasks completed. Moreover, experiment 2 participants completed additional elements that were not controlled for (e.g. through the use of a "filler" task) in experiment 1, which creates an additional experimental confound [105]. It is also not possible to attribute the increase in accuracy from 48.2% to 59.0% to any one intervention, since experiment 2 was different to experiment 1 in two ways: participants received feedback as well as a short tutorial with examples of rendering artefacts that could be used to differentiate real from fake. Since neither of these interventions were present in experiment 1 (which could be seen to act as a control condition), the experimental setup does not allow us to attribute this 10% difference in accuracy to either intervention in isolation. To isolate the effect of these two interventions would require that they were employed individually as well as in combination, and that participants be randomly allocated to experimental conditions (see above).

In the final experiment carried out as part of this study, participants were asked to report on the "trustworthiness" of deepfake and real face images. In this experiment, participants reported deepfake images to be "trustworthy" significantly more than real images: 4.82 as against 4.48 on a scale of 1 ("very untrustworthy") to 7 ("very trustworthy"). The study authors hypothesised that this difference was due to the deepfake faces looking more like average faces, which some research indicates tend to be more trustworthy [106]. However, it should be noted that while the effect (of 0.34) was statistically significant it is a tiny effect. As such, it is unclear whether this would have any practical significance in the real world.

Other details regarding the design of the study also warrant consideration. Perhaps most notable is that the image dataset was not a representative (i.e. random) selection of the FFHQ real face dataset, or of the StyleGAN2 algorithm as trained on the FFHQ dataset. The study authors curated the dataset in several ways. The real images were selected such that they each had a uniform background, an unobstructed face, and had no other "extraneous cues" such as visible brand logos or writing. The deepfake image dataset was also selected such that images did not have any "obvious rendering artefacts" and had a uniform background. This curation element raises the question of what comprises an "obvious rendering artefact", given that the treatment group were given a short tutorial on "rendering artefacts" that help to identify an image as deepfake. The difference between a "rendering artefact" and an "obvious rendering artefact" is not, in this case, obvious. If no difference was indeed intended, then participants were given a tutorial on how to look for "obvious rendering artefacts" which had been removed from the image deepfake dataset that they were then tested on. Finally, as with the first study discussed, cognitive overload was a potential issue in this study. To explain, participants were tasked with labelling 128 image items, which is a lengthy and repetitive experimental task.

In the third study to evaluate human accuracy at detecting image deepfakes from authentic images [107], participants were given a pair of images, one real and one deepfake, and asked which of the two was the deepfake image. In the main experiment, 176 participants recruited from Amazon Mechanical Turk answered correctly only 49.1% of the time. The study authors hypothesized that participants might have been considering elements other than features of the faces they were shown to decide whether an image was a deepfake or authentic. For this reason, they repeated the experiment but using images for which the backgrounds were replaced with a solid black colour, or where the images had a synthetic randomly-directed sunlight effect overlaid upon them. The accuracy for each of these two additional experiments was 49.7% (N=174, and N=172).

The manipulation of the backgrounds used was an interesting innovation that could be implemented at scale if it had proved useful. However, in methodological terms, this study also has notable limitations. First, with regard to how the authors sourced the image stimuli: the authentic image stimuli were taken from a dataset [108] consisting of photographs collected from 2002 through 2012. The dataset in question (which is not publicly available) holds a variety of types of photograph and other media, so the study authors will likely have had to manually select suitable images. In doing so, they will have curated the dataset of authentic image stimuli. What criteria were used to do this are unclear. In addition, participants were each given 50 questions to answer, which may have affected the study findings by introducing an element of cognitive overload.

There is also the question of ecological validity; that is, to what extent the experimental set-up emulates a real-world context. In the context of the current work, this is important as we are concerned with real-world implications. In the Shen et al. study [107], participants were given two images at once, and told that one of them was a deepfake and that the other was real. It is hard to imagine this scenario, a forced choice between two options, occurring in real-life misuses of the technology (e.g. for a dating scam).

In summary, the existing three studies represent important work that provides insight into the human detection of image deepfakes. However, as discussed, each of these studies has one or more methodological issues that limit the conclusions that can be drawn. Moreover, none of the studies conducted so far have used an experimental procedure that enables the testing of whether training can improve the human detection of deepfake images.

### 1.4. Human Detection Ability Studies: Video Deepfakes

Several other studies have examined human detection ability with respect to video deepfakes. Although this difference in modality will have complex effects on the human detectability of the deepfakes, the findings from these may shine light on human image deepfake detection. However, the findings from the eight studies available [73,98,109–114] do not lead to a consensus: there is a surprisingly large range in human accuracy in the labelling of deepfake stimuli across studies from 23% [109] to 87% [110]. Human accuracy at correctly labelling real, non-deepfake videos shows a much smaller range (from 75% [98] to 88% [110]), and seems to indicate that humans are relatively reliable at labelling real video stimuli even in the presence of deepfake stimuli.

However, too much weight should not be put on the findings from these studies either. Most were quite vague about the study design and would be difficult if not impossible to replicate. Many suffered from having small sample sizes (across the eight studies, $N$ = 14, 20, 30, 55, 204, 204, 210, 301), and several of the studies spread participants across multiple conditions (such as varying the time length, compression, or resolution of the video stimuli, or varying the deepfake generation method, or even varying the demographic category of the face in the image stimuli), meaning that the number of participants per condition was small, even if the overall sample size was not. The low sample sizes employed in most studies mean that the confidence intervals associated with the reported estimates would be large, making them hard to interpret in a meaningful way. Moreover, we note that in only one of the studies was an a-priori statistical power analysis reported as being conducted to ascertain the sample size required to detect effects should they exist [112]. Furthermore, descriptive statistics were often absent or hard to extract from the study findings [61], and estimates of uncertainty (standard errors and $p$-values) were mostly omitted. On a different note, in four of the studies, participants appear to have been recruited from computer science courses, sometimes even from those at the authors' institution [73,110,113,114], meaning that the study findings are far from generalisable. Of the remaining four studies, one obtained participants from Amazon Mechanical Turk [109], two from Prolific [98,112], while two others do not state how participants were recruited [61,111]. In defence of these studies (and study [61]), for all but three [98,110,112] the focus was on the technical challenge of creating either a deepfake generation method or a computational deepfake detection method. As such, estimating human accuracy was presumably a secondary element.

An additional issue in this literature is the deepfake generation method chosen for testing. In half of the studies reviewed, the method tested was also created by the study authors [73,109,111,114], which introduces a potential conflict of interest. In all of these cases and also the three papers that used the DFDC Kaggle dataset [98,112,113], the deepfake generation method used was also closed-source. This means that they are not open to inspection nor necessarily representative of the quality of open-source deepfake methods which - being open-source and widely available - will comprise most of the threats that deepfakes pose, at least for the present. Furthermore, in the case of the [112] study the stimuli were cherry-picked to be the most convincing deepfake stimuli from the DFDC Kaggle dataset. This means that the stimuli used in that experiment were not representative of even the closed-source deepfake generation methods used to generate the DFDC Kaggle dataset.

One study [112] avoids most of these errors but has another methodological pitfall. The study had two treatment groups and one control group. Prior to testing, one of the treatment groups was tasked with the comprehension of over 400 words (from [51]) to inform them of the dangers of deepfakes, but the other two groups were not given a filler task, which introduces an experimental confound. However, since this is the literature's least problematic study its findings are important to discuss. The task was labelling a video stimulus as authentic or fake; one treatment group was given an additional monetary incentive to perform well, while the other treatment group read a text excerpt detailing the potentially harmful consequences of deepfakes. Neither treatment affected self-reported motivation levels and nor did it increase accuracy on the experimental task. Participants had a tendency to label stimuli as "authentic", doing so 67.43% of the time (though they were informed at the start of the experiment that the likelihood of stimulus authenticity was 50%). Participant accuracy across conditions was 57.6%, which significantly exceeded chance performance but is self-evidently a poor level of accuracy. Despite this, and despite accuracy ranging from 46.7% to 77.6% across the videos used, self-reported levels of confidence levels were consistently high (remaining within the range 73.7% to 82.5%, with 100% indicating complete confidence).

The Groh et al. [98] study is also worthy of discussion, since it recruited 304 paid participants via the Prolific platform[3]. In this study, participants completed twenty single-item fake-or-real tests; once participants had submitted a fake/real label for each item, they were told which label had been given to the item by a computational deepfake-detection algorithm, and they were then asked to submit a new label (changing it from their initial answer if they wished). Participants' mean accuracy was 66%, which exceeds chance performance but is not particularly high. In terms of the effect of training on accuracy, the design of the study consisted of a treatment condition only, meaning that there was no estimate of participants performance in the absence of intervention.

With such a limited evidence base regarding static images, only a single study with a reasonably rigorous methodology for deepfake videos [112], and no studies that have tested the effects of training with an appropriate experimental design, there is a pressing need for studies that employ experimental procedures that use appropriate samples, sample sizes, and deepfake generation methods to examine human deepfake detection accuracy. We do that here.

### 1.5. StyleGAN2

A variety of deepfake generation methods have been used in previous studies; however, one method in particular currently stands out as appropriate and relevant for an assessment of human accuracy in the detection of deepfakes – StyleGAN2 [16]. StyleGAN2 can produce stationary deepfake images in the "style" of practically any content [115] (such as animals, cars, room interiors), but has demonstrated particular proficiency in producing deepfake images of human faces. It is able to create high-resolution (1024x1024 pixels) deepfake images and there exists publicly-available source code that is currently implemented on a server which outputs one deepfake instance every two seconds to a public-facing website [116,117]. This means that at present no technical expertise or resources are required to obtain these deepfakes at scale. Once they have been obtained, they can be used for a range of purposes, including criminal ones. For instance, the images of human faces could be used as profile pictures for bot or scam accounts on social media, including dating websites or apps. Doing so could have the effect of raising the perceived reputability of such accounts [118,119], and so existing problems of fraud and catfishing may be exacerbated by the incorporation of StyleGAN2 deepfakes. The technology might also be exploited to open fraudulent bank accounts through so-called "challenger" banks that lack physical branches and rely on uploaded images (rather than in-person visits) to authenticate people's identity (though note that KYC checks mitigate this risk [120]). The (mis)use of deepfakes as profile pictures has already been reported in the wild [121–123] and consequently there is an element of urgency associated with assessing human accuracy in detecting deepfake instances generated by this method.

---

[3] This study also reports findings from responses given by separate groups of participants who organically found the experiment's website and submitted responses to one or more images (n = 6390, 9188). We do not evaluate the findings from these participants' responses, for several reasons. These participants were not paid so were not financially motivated to perform well; additionally, there was no lower bound set on the number of responses a participant had to submit in order for their data to be included in the findings. These participants submitted an average of 4.86 responses to questions in the first experiment (no paid Prolific participants were recruited for this experiment, so we do not mention it in our discussion), and (separate non-Prolific participants) submitted an average of 6.71 responses in the second experiment (which we do discuss). In the case of the second experiment, reported details indicate a high level of variance within the number of questions answered per participant: only 10% of these participants answered 17 or more questions. These reasons, as well as the unknown nature of these online participants (the reported methodology does not indicate any implemented means of stopping the same participant from retaking the study, for instance) we avoid discussing these participants above. However, we do note here that the authors of that study indicate that the findings for these participants echo the findings of the Prolific participants, although again the nature of the participant groups and the degree of their participation differs to an extent that equation of their results is not a simple matter.

However, while the quality of the images and the speed with which they can be produced are impressive, StyleGAN2 deepfake images do commonly include visual flaws that are potentially obvious to a human observer [124]. The presence of these imperfections make StyleGAN2 a yet better test case – if participants cannot detect this type of deepfake, they will surely struggle with more robust methods or future improvements to this technology. In the next section, we report the findings of our study, which uses StyleGAN2 (as trained on the Flickr FFHQ dataset [100]) to generate the images tested.

### 1.6. Research Questions

Our Research Questions were as follows:

RQ1 – Are participants able to differentiate between deepfake and images of real people above chance levels?

RQ2 – Do simple interventions improve participants' deepfake detection accuracy?

RQ3 – Does a participant's self-reported level of confidence in their answer align with their accuracy at detecting deepfakes?

### 1.7. Contributions

This study offers an investigation into the ability of humans to detect deepfake images from similar authentic images, and into the methods by which they do so. It differs from the existing published literature in the following ways:

1) This study uses (and reports) an a-priori statistical power analysis to ensure that the sample sizes used would be sufficient to detect differences between conditions should they exist. The experiment is constructed to avoid cognitive overload (each participant examines only 20 images), the control condition features a filler task (thus avoiding an experimental confound), all participants are randomly allocated from the same participant pool (avoiding allocation error), and the real and deepfake datasets are each from one data source (FFHQ and StyleGAN:FFHQ respectively). The survey for the study was prepared on a custom-built Django webapp which uses JavaScript to ensure randomization at every relevant point, for instance randomization of the order in which the 20 image stimuli in the familiarization intervention appear.

2) The study has a strong focus on ecological validity and relevance to the wider context of the criminal misuse (e.g. in cases of fraud, such as dating scams) of the technology and policy implications. As such, the experiment was designed to emulate a realistic scenario. For example, participants could have been asked to compare two images – one real and one fake – and to indicate which was which. This is the approach taken in the Shen et al. study [107]. However, it is hard to imagine a real-world scenario in which people would need to do this. A realistic situation would be when an individual encounters a single image (e.g. posted by a dating scam fraudster) and is required to determine if that is real or not. For this reason, we avoid the use of a two-alternative forced-choice task and asked participants to judge each image on its own (as they would in the real world). Moreover, the deepfake data source chosen is StyleGAN2 which, at the time of the experiment, represented the cutting edge in image deepfake generation. Importantly, instances of StyleGAN2 (indeed, of StyleGAN2:FFHQ) are publicly available at scale and so could be used by anyone [116]. The real image data source chosen is FFHQ, which is a dataset of images of regular individuals' faces as pulled from a social media photograph-sharing platform [100], thus being fairly representative of the types of image that a person might see on the internet (as opposed to being representative of a laboratory setting). The two data sources are not only very similar, but the deepfake source has been trained on the real image source, such that the image stimuli for the experiment in this study are as similar as possible to one another whilst still being strongly representative of real and deepfake images as they might be found in the wild.

3) The study delves deeper into participant decision making. For example, participants were asked how confident they were about the correctness of each of their responses, and correlations were computed to examine the relationship between their confidence and accuracy. Participants were asked to supply the reasoning behind each of their responses, both by providing an answer to an open question and by indicating which part(s) of the image their responses apply to. These verbal and visual reasonings are inspected for each response and analysed to establish whether participants used reasoning that matches any of the elements of the experiment's advice intervention or not. After this coding and categorisation, the data is split to establish the extent to which participants' use of the advice intervention's elements was applied to deepfake images (correct application) or real images (incorrect application). Results are also inspected on a per-image level, which allows an analysis of whether some images are universally more real-seeming than others.

## 2. Materials and Methods

### 2.1. Participants

An a-priori statistical power analysis was conducted to estimate the sample size necessary to detect a "moderate" (Cohen's $f$ = 0.25) effect size across experimental conditions (see below). The power analysis was conducted for a one-way ANOVA with four groups, an alpha significance level of $\alpha$ = 0.05, and statistical power of 0.95. The resulting overall sample size estimate was 280 participants.

Participants were recruited using the Prolific online platform [125] and received a payment of £6/hr. To increase motivation, they were informed that they could earn a 50% bonus payment if their performance was in the top 50% of all participants. Participants were informed of this bonus payment immediately before the experimental task began, and were reminded each time they had to judge an image. Bonus payments worth 50% of the original payment were allocated after the experiment was complete.

Table 1 provides summary statistics for participants allocated to the four conditions. The variation in sample sizes reflects the use of simple random allocation. Participants had a mean age of 26.24 and 39.78% were female. Of the 273 participants who volunteered their nationality information, 15 (5.5%) were UK nationals, while 229 (84%) were (non-UK) European nationals. 231 participants volunteered information about their first language, which for 21 participants (9%) was English and for 205 participants (89%) was a European language. However, this diversity in nationality had no impact on comprehension of the experimental task or any elements of the survey: where textual input was required of participants full comprehension was demonstrated by all. There were no differences in any of these characteristics across conditions.

*Table 1: Participant Characteristics Across Conditions*

|                      | Sample Size | Age (SD)      | Female (%) |
|----------------------|-------------|---------------|------------|
| Control              | 72          | 25.03 (8.67)  | 30.56      |
| Familiarisation      | 65          | 29.35 (10.12) | 48.44      |
| One-Time Advice      | 82          | 25.48 (7.26)  | 41.25      |
| Advice with Reminders| 61          | 25.27 (7.85)  | 39.66      |
| Overall              | 280         | 26.24 (8.62)  | 39.78      |

### 2.2. Design

The study comprised four conditions to which participants were randomly assigned – a baseline control condition and three "experimental" conditions intended to improve participants performance on the experimental task. The interventions were designed to be simple, scalable, and suitable for widespread use. The first experimental condition was a familiarization intervention for which participants were shown 20 examples of deepfake images and informed that that was what they were. They were asked to spend a few moments looking at each image, and told that this was their only chance to familiarize themselves with the images before they would complete the experimental task. The same 20 images, none of which were used for the experimental task, were used for each participant but their order was randomised each time.

In the second and third experimental conditions, participants were shown a list of ten "tell-tale features" (see Figure 1) that deepfake images of this kind commonly contain and that may be used to distinguish them from non-deepfake images. For each item, they were provided with a brief description of the features and two deepfake images that illustrated them. For the second condition, the advice was shown only before the experiment started. For the third, an additional reminder of the names of each tell-tale feature was displayed beneath each image throughout the experiment (see Figure 1).

To encourage participants to pay close attention, they were told that they would be quizzed on their memory for these features later. They were also informed that these tell-tale signs would not necessarily feature in each AI-generated image, that they were not the only features that could be used to distinguish the AI-generated images, and that it was important that participants applied common sense when making decisions.

| Abnormalities | Asymmetries | Background |
|---|---|---|
| Accessories don't make sense | Earrings | Colours bleed in from background |
| Strange clothing fabric | Glasses | Impossible location |
| Abnormal clothing structure | Facial hair | Warped companion faces |
|  |  | Distorted text/patterns |

*Figure 1: Screenshot of the Textual Advice Reminders that participants within the Advice with Reminders condition received*

To ensure comparability across conditions, participants in the Control condition completed a filler task to replace the Familiarization or Advice tasks that those allocated to the other three conditions received. This filler task was designed to have a neutral impact on performance. Specifically, participants were told that we were interested in developing appropriate greeting and thank you messages for subsequent studies. They were then shown a series of greeting messages and asked to rank them according to how engaging, professional, and uplifting they perceived them to be. Participants were also asked to give their own variant. This process was then repeated but for a series of "thank you" messages.

The key aspect of the experimental task was a binary question asking participants whether they believed a given image was a deepfake or not. Of course, differentiating between deepfake and genuine images is an absolute task – people are either right or wrong. In the real world, when people make the types of decisions that deepfakes may be intended to affect, decision-making is unlikely to be quite so binary. People may have their doubts about an image, which may in turn affect the decisions they make. Consequently, as well as assessing participants' accuracy for an experimental task, we also assessed the confidence participants expressed regarding their decisions, which enabled us to examine whether this is (for example) uniformly low for deepfake images, whether they are more accurate for images that they are most confident about, and whether those who are the most confident are also the most accurate.

### 2.3. Materials

One aim of the study was to determine whether each participant's performance (labelling a stimulus as AI-generated or not) exceeded chance. As such, it was important to require each participant to complete a sufficient number of trials so that this could be detected individually as well as in the aggregate. After computing the standard error of a proportion for different numbers of trials, we reasoned that 20 trials would provide a good trade-off between providing sufficient precision while minimising the risk of cognitive overload.

Images were selected from a pool of 50 real and 50 deepfake images. The fifty images of real human faces were drawn at random from images in the Flickr-Faces-HQ Dataset (FFHQ) [100], a dataset of high-resolution images sourced from Flickr that only contains photos of human faces. In this dataset, each photo underwent automated alignment and cropping such that the human face takes up the majority of the image. As such, the dataset is a collection of photos of real people in which their faces can be seen in great detail.

The Deepfake images were generated by the StyleGAN 2 deep learning algorithm [64], trained using 70,000 images from the FFHQ dataset discussed above. All deepfake images were collected without any element of curation, and so represent random output of the StyleGAN 2 algorithm.

## 2.4. Procedure

The experiment was implemented using a web application which was written in Django (available at https://github.com/sergibot/FakeFacesSurvey) and hosted on a secure server in The Netherlands. Participants first read about the study aims and the conceptual distinction between images of real human faces and AI-generated equivalents. Instructions were then provided to explain how they should complete the experimental task, and an example provided of how the experimental task webpage should be completed. They were then provided with details about how their data would be stored (anonymously) and told that they had the right to withdraw from the experiment at any point without giving a reason. They were subsequently asked to complete a series of checkboxes to provide informed consent and provided with the contact details of the first author should they have any questions about the study prior to participating. On providing consent, they were randomly allocated to one of the four study conditions.

Participants then viewed the intervention content of the condition to which they had been allocated and completed the Experimental Task. Participants were shown one image at a time and no participant was shown the same image twice. Images were consistently displayed at a size of 720x720 pixels. For each image, participants were asked to label the image as "AI-generated" or "real" and to rate their confidence in their answer on a 9-point scale (from "No confidence at all" to "Complete confidence"). They were additionally asked to answer a free-text question about the reasoning behind their decision and to click on a ten-by-ten grid (see Figure 2) that was subsequently superimposed over the image to highlight which parts of the image influenced their decision. Finally, all participants were shown a "Debrief" page which explained the contents of the different interventions tested.

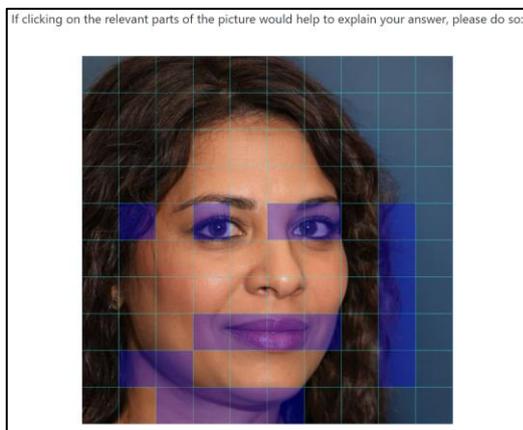
Figure 2: Example of Highlighting Explanatory Regions of Image during Experiment

## 2.5. Ethics

The experiment received ethical exemption from the Departmental Ethics Committee.

## 3. Results

### 3.1. Overall Accuracy

Table 2 shows the mean number of items participants correctly identified as real or deepfake images overall. On average, participants were correct about 60% of the time. One-sample $t$-tests confirmed that accuracy was significantly greater than chance (50%) for those in the Control condition ($t(71) = 5.81$, $p < 0.001$, Cohen's $d = 0.685$), the Familiarization condition ($t(64) = 6.02$, $p < 0.001$, Cohen's $d = 0.746$), the One-Time Advice condition ($t(81) = 9.29$, $p < 0.001$, Cohen's $d = 1.03$), and the Advice with Reminders condition ($t(60) = 7.04$, $p < 0.001$, Cohen's $d = 0.901$).

Table 2: Deepfake Detection Accuracy Results (group sample sizes differ due to the use of a simple random allocation strategy)

|  | Sample Size | Mean (%) | Standard Deviation (%) |
| --- | --- | --- | --- |
| Control | 72 | 59.65 | 14.10 |
| Familiarisation | 65 | 60.54 | 14.12 |
| One-Time Advice | 82 | 64.15 | 13.78 |
| Advice with Reminders | 61 | 64.26 | 15.83 |
| Overall | 280 | 62.18 | 14.48 |

While performance was marginally better for those in the experimental conditions, a one-way Analysis of Variance ($F(3,276) = 1.95$, $p > 0.1$ $ns$) suggested that there were no statistically significant differences between groups. Simply put, for overall performance, none of the intervention conditions led to a significant improvement in participants' accuracy at detecting deepfake from non-deepfake images.

Table 3 provides a more detailed breakdown, showing accuracy for real and deepfake images separately (measured as the percentage of the total number of each stimuli that was correctly labelled). A 4 (condition) X 2 (real versus deepfake image type) repeated measures ANOVA showed no significant main effects ($ps > 0.15$ $ns$), but a statistically significant interaction ($F(3,276) = 6.96$, $p < 0.005$). In particular, those assigned to the experimental conditions were more likely to correctly identify deepfake images, and accuracy increased with the "intensity" of intervention. However, this improvement was – to varying degrees – offset by a decrease in accuracy at labelling non-deepfake images. This was to the extent that overall accuracy did not significantly differ from that for those in the Control condition.

Table 3: Accuracy Split by Condition and Image Truth (Real/Fake) (standard deviations in parentheses)

|  | Sample Size | Accuracy for Real Images (%) | Accuracy for Deepfake Images (%) |
| --- | --- | --- | --- |
| Control | 72 | 68.40 (21.90) | 51.75 (24.87) |
| Familiarisation | 65 | 59.63 (21.42) | 62.01 (23.52) |
| One-Time Advice | 82 | 65.77 (19.84) | 62.25 (26.70) |
| Advice with Reminders | 61 | 60.49 (24.13) | 69.10 (27.01) |
| Overall | 280 | 63.87 (21.90) | 60.99 (26.18) |

### 3.2. What influenced decision making?

As well as indicating whether each image was "AI-generated" or "real", participants were asked to provide a "free text" explanation as to the reasoning behind their decisions and to click on locations within the images that informed their choices. Here, the free text data was manually coded using NVIVO [126,127] to count the frequency with which participants used any of the ten "tell-tale signs" provided in the Advice intervention condition. For instance, if a participant reported that the "background was impossible", this was coded to the "Background: Impossible Location" item. Account was also taken of where – within an image – participants clicked. For example, if they clicked on a part of the image that clearly included a tell-tale sign, this was coded accordingly.

Figure 3 shows the proportion of responses that contained one or more of the advice items. Only items labelled as deepfakes are considered and we differentiate between those that were correctly labelled as such and those that were not (i.e. non-deepfake images labelled as deepfakes).

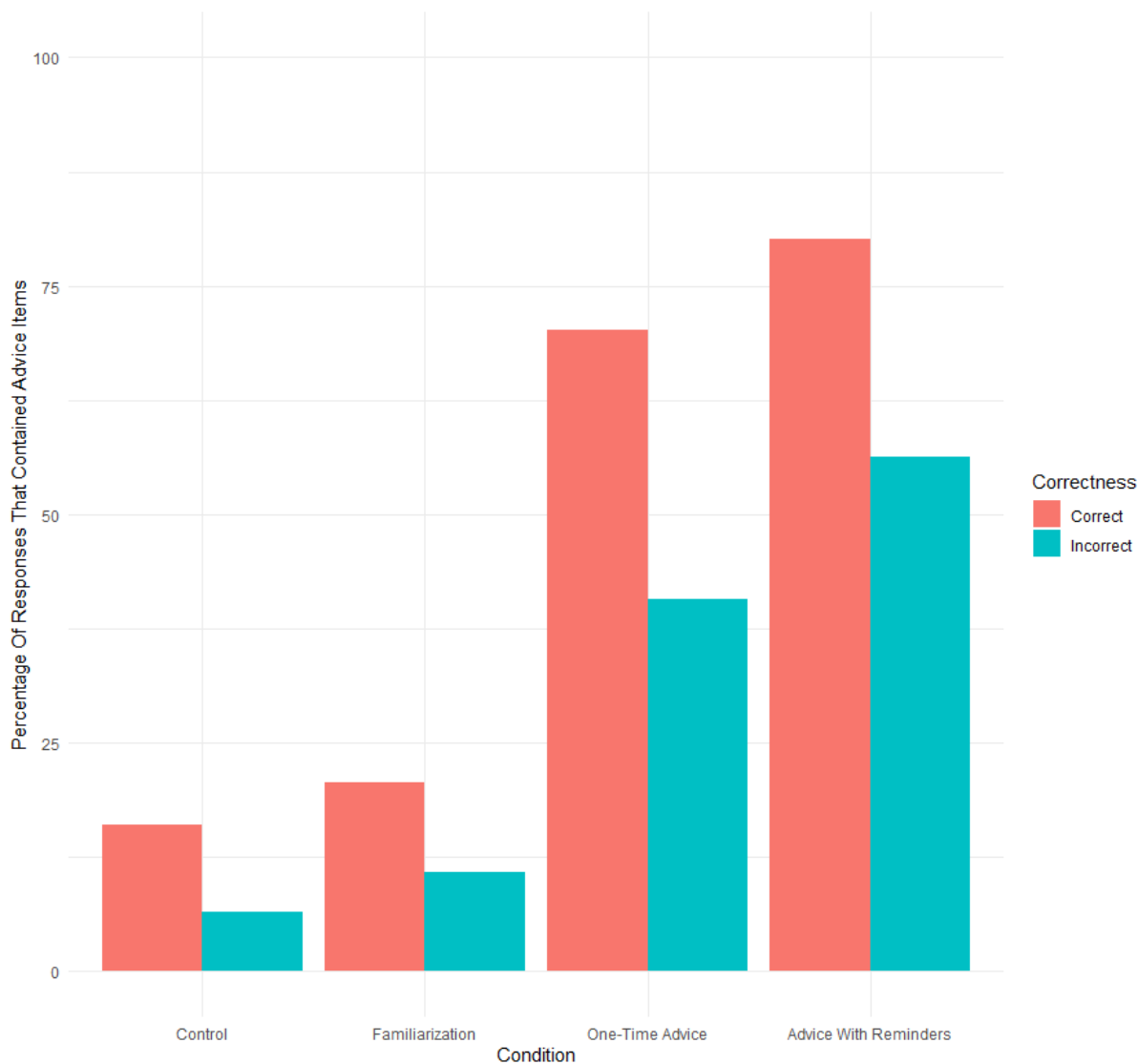

*Figure 3: Use of "tell-tale signs" in participants' reasoning behind response decisions, by decision correctness (where participants' responses contained "Deepfake" as the selected label)*

Figure 3 suggests that those assigned to the advice conditions were more likely to report using features included in the advice provided to identify deepfake images than were those in the control or familiarisation conditions. Across conditions, the use of these features was more likely to be associated with correct attributions, but it is evident that – in the aggregate across advice items – their use was suboptimal as participants frequently also used these as discriminating features when they labelled real images as deepfakes. An unanticipated consequence of the intervention then appears to be that it increased the number of non-deepfake stimuli that participants incorrectly labelled as "deepfakes". This may explain why performance improved for deepfake images but decreased for non-deepfake images – participants allocated to the advice conditions appear to have simply labelled more items as deepfakes.

To examine this more explicitly, we examined the number of deepfake labels that participants applied. In this experiment, the images that participants were shown were drawn at random from a large pool of image stimuli (without replacement) which had a 1:1 ratio of deepfake image stimuli to non-deepfake image stimuli (which resulted in a mean of 10 deepfake image stimuli seen per participant, *SD* = 2.01). Participants were not made aware of this 1:1 ratio, the randomness of the draw, or how many faces they would see so that they could not use such information to guide their decision making.

If participants assumed that images had a 50-50 chance of being a deepfake, this would result in them labelling around 10 images as fakes. However, if the number of deepfake images was unbalanced across conditions this could distort the results. Table *4* shows the number of deepfake stimuli observed across conditions, along with the mean number of deepfake labels applied. It is evident that participants in the Control condition saw the same number of deepfake images as everyone else ($F(3,276) = 0.41$, $p > 0.7$ *ns*), but labelled significantly fewer pictures as such ($F(3,276) = 5.06$, $p < 0.005$). This was the only significant difference observed and shows that those allocated to the advice or familiarisation conditions did label more items as deepfakes.

Table 4: Number of Deepfake Labels Applied (standard deviations in parentheses)

|  | Sample Size | Number of Deepfake Stimuli Actually Seen in Questions | Number of Deepfake Labels Applied in Questions (Regardless of Correctness) |
| --- | --- | --- | --- |
| Control | 72 | 10.03 (1.85) | 8.43 (3.72)* |
| Familiarisation | 65 | 10.18 (2.08) | 10.23 (3.39) |
| One-Time Advice | 82 | 9.82 (2.16) | 9.72 (3.91) |
| Advice with Reminders | 61 | 9.98 (1.92) | 10.87 (4.05) |

*$p<0.001$

In addition to looking at the association between the use of the tell-tale advice items in the aggregate, we examined their efficacy individually. Figure 4 shows how frequently each "tell-tale sign" was identified when participants labelled an image as a deepfake. It is clear that none were used every time, but that some were used more frequently than others. In particular, the "Accessories Don't Make Sense" item was used most often, but was used incorrectly almost as many times as it was used correctly. In contrast, the "Strange Clothing Fabric" and "Asymmetric Earrings" signs were used less often but were associated with correct answers most of the time. Put simply, some items were more useful than others.

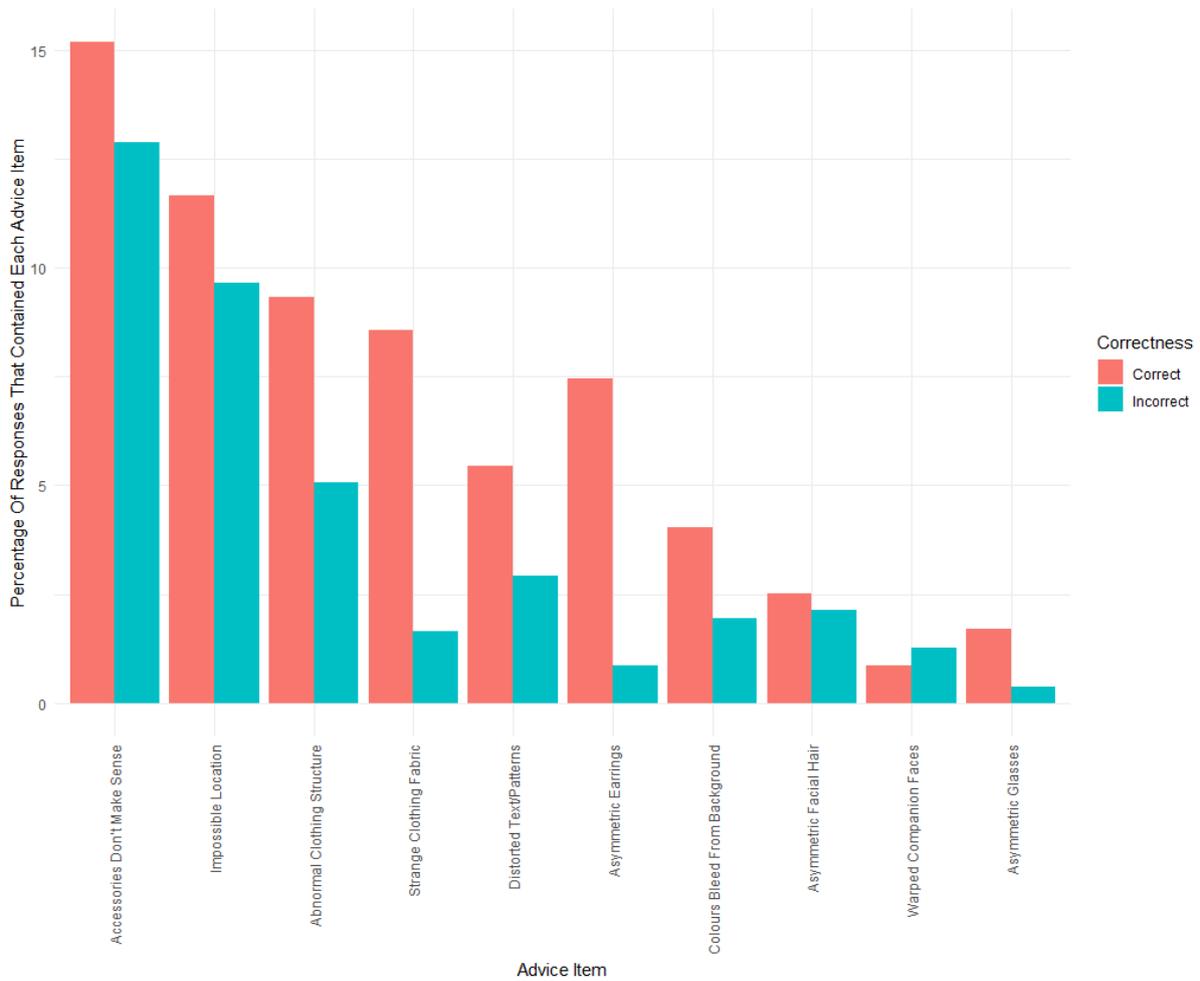

*Figure 4: Use of each "tell-tale sign" in participants' reasoning when choosing "deepfake" label (with percentages showing mean accuracy across conditions)*

### 3.3. Confidence

The second column of Table 5 shows that participants' mean confidence in their decisions was typically high. A one-way ANOVA revealed that the differences observed across conditions were statistically significant ($F_{(3,276)} = 3.72$, $p < 0.05$), but as is visible from Table 5, these differences were generally small and hence follow-up $t$-tests (see SI) are not discussed here.

Table 5: Mean Confidence Results (standard deviations in parentheses)

|  | Sample Size | Overall Confidence | Labelled Real | Labelled Deepfake | Image Real | Image Deepfake |
|---|---|---|---|---|---|---|
| Control | 72 | 6.77 (1.41) | 7.15 (1.42) | 5.92 (1.52) | 6.83 (1.52) | 6.73 (1.48) |
| Familiaris- ation | 65 | 6.47 (1.56) | 6.57 (1.72) | 6.30 (1.63) | 6.42 (1.69) | 6.52 (1.55) |
| One-Time Advice | 82 | 7.18 (1.08) | 7.18 (1.67) | 6.99 (1.78) | 7.11 (1.22) | 7.26 (1.08) |
| Advice with Reminders | 61 | 6.80 (1.12) | 6.91 (1.29) | 6.52 (1.30) | 6.63 (1.32) | 7.05 (1.17) |
| Overall | 280 | 6.83 (1.32) | 6.97 (1.42) | 6.46 (1.45) | 6.77 (1.45) | 6.90 (1.35) |

Splitting the responses by the label that a participant applied (deepfake or not) suggests differences between conditions and participant responses. A 4 (condition) X 2 (labelled real versus labelled deepfake) repeated measures ANOVA revealed that there was a significant main effect of condition ($F(3, 276) = 3.88$, $p < 0.01$), a significant difference for images labelled fake or not ($F(1, 276) = 29.03$, $p < 0.0001$), and a statistically significant interaction ($F(3, 276) = 8.25$, $p < 0.0001$). Again, while the differences were reliable, they were small and hence we do not report follow-up t-tests here but provide these in the SI. Put simply, the condition a participant was in had a slight impact on the confidence they had when they applied a "deepfake" label as opposed to a "real" label.

Comparing responses by the type of image (i.e. whether the image being labelled is actually a deepfake or real), a 4 (condition) X 2 (image type) ANOVA with repeated measures on the second factor revealed a significant main effect of condition ($F(3,276) = 3.68$, $p < 0.05$), a significant difference for image type ($F(1,276) = 6.29$, $p < 0.05$) and a significant interaction ($F(3,276) = 3.42$, $p < 0.05$). Follow-up $t$-tests reveal statistically significant differences between conditions but differences were again small. Again, the important point here is that the condition a participant was allocated to had a slight impact on the confidence that they had when applying a label to an image that was indeed a "deepfake" as opposed to a "real" image. However, the differences were small.

### 3.4. Confidence-Accuracy Correlation

There are two ways in which participants' confidence may be correlated with their deepfake detection abilities. First, we calculated a within-subject correlation between the confidence and accuracy of responses for each participant across the twenty questions asked. Goodman-Kruskal Wallace Gamma Correlations were used for this purpose and showed that while the mean correlation for those in the control condition ($M\gamma = 0.06$, $SD = 0.39$) was not significantly different from zero ($t(69) = 1.27$, $p > 0.2$), the mean correlations for those in the Familiarization ($M\gamma = 0.13$, $SD = 0.35$), One-Time Advice condition ($M\gamma = 0.21$, $SD = 0.46$), and Advice with Reminders conditions ($M\gamma = 0.14$, $SD = 0.50$) were (respectively: $t(63) = 3.08$, $p < 0.005$; $t(80) = 4.11$, $p < 0.001$; $t(58) = 2.17$, $p < 0.05$). However, the coefficients were uniformly low. In other words, across any given participant's responses, there was not very much correlation between the confidence they had in their answer, and the accuracy of their answers.

Second, we computed between-subjects Pearson's correlations to examine whether those who were overall more confident were also overall more accurate. None of the correlations were statistically significant (Control: $r(70) = 0.04$, $p > 0.7$ ns; Familiarization: $r(63) = -0.02$, $p > 0.9$ ns; One-Time Advice: $r(80) = -0.51$, $p > 0.7$ ns; Advice With Reminders: $r(59) = 0.02$, $p > 0.9$ ns). That is, participants who were more confident were not necessarily more accurate.

### 3.5. Are all images equal? Per-Image Analysis

Figure 5 shows the mean accuracy rates per image, aggregated across conditions. It shows that accuracy was not uniform across images. For some (real or fake), mean accuracy was well above 50%, whereas for others (about 20% of images), participants attained a mean accuracy below this threshold. Figure 5 also shows the mean confidence ratings calculated across participants. While there appears to be no association, a between-item Pearson's correlation revealed a statistically significant correlation between mean confidence and mean accuracy ($r(98) = 0.33$, $p < 0.001$). The effect is, however, only moderate and the differences in mean confidence vary within a very small range of values (5.82 to 8.10).

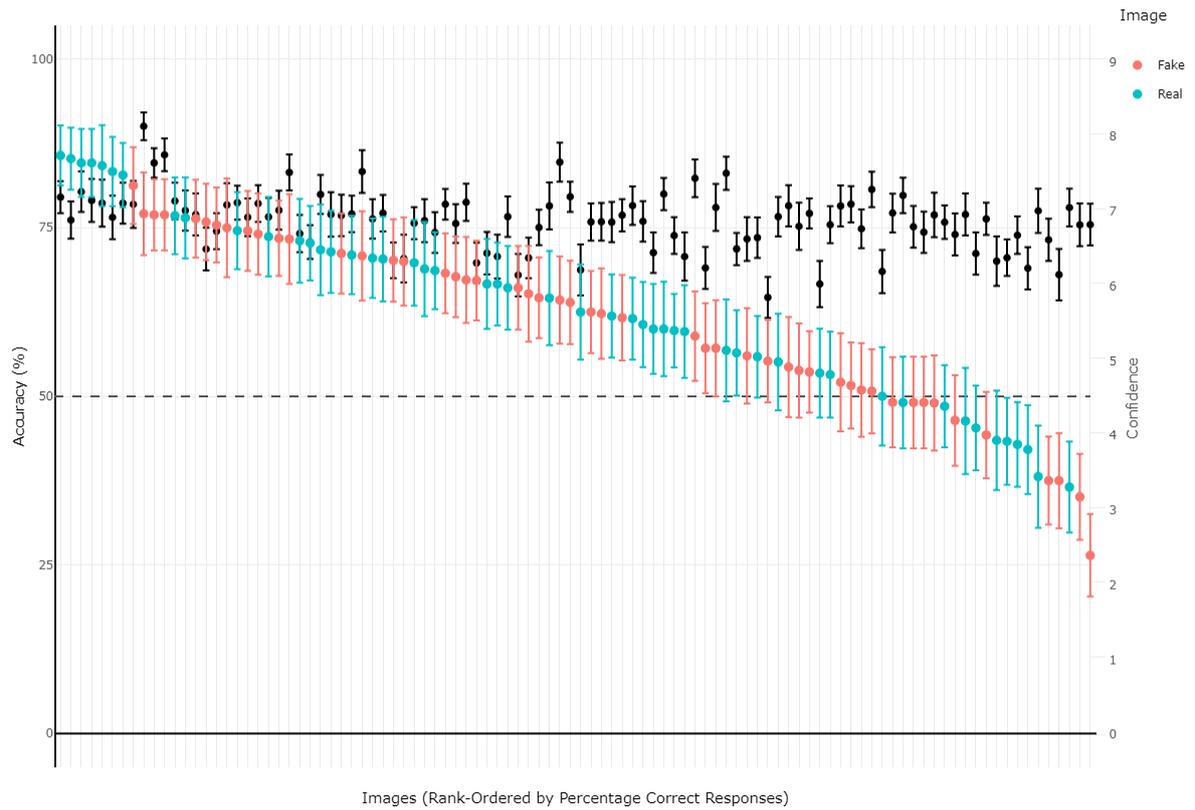

*Figure 5: Results Split by Image (light colours = Accuracy, black = Confidence)*

## 4. Discussion

### 4.1. Participant Performance

Algorithms to generate "deepfakes" are increasing in sophistication and availability. While there are positive use cases, many potential criminal applications also exist. Previous research has examined people's ability to detect deepfakes but much of the research is subject to important limitations (see above) and has used a variety of algorithms to generate deepfakes, which are of variable quality. Here, we test human ability to detect deepfakes generated by StyleGAN2, the output from which is freely available. We find that although deepfake detection accuracy was significantly better than chance, mean accuracy ranged from only 60% to 64% across conditions suggesting that humans are not very good at detecting deepfake images generated by StyleGAN2. Moreover, we do not find that there was a significant improvement in overall accuracy for participants who completed a familiarization exercise, or for those who were provided with explicit advice as to how to identify deepfakes.

Three studies previously measured the deepfake detection abilities of participants for static images. The first [61] used deepfake images representative of the earlier version of StyleGAN [99] as trained on the FFHQ [100] and CelebA-HQ [101] datasets, and representative of the PGGAN algorithm as trained on the FFHQ dataset. Accuracy achieved (across $N$=20 participants) ranged from 63.9% to 79.13%, with accuracy being the lowest for StyleGAN1 as trained on FFHQ. While the sample sized used in that study precludes firm conclusions, the level of accuracy for StyleGAN1 is within the range of participant accuracies found the current study. The second study [104] used deepfake images from StyleGAN2 also trained on FFHQ, but excluded images with "obvious rendering artefacts" from the dataset, thereby making the labelling task more difficult for participants. Accuracy varied across experiments from 48.2% ($N$=315 participants) to 59.0% ($N$=219 participants) – results which are relatively consistent with our own.

The third study [107] used equal numbers of deepfake images from StyleGAN2 as trained on FFHQ, and from SREFI [128] as trained on an academically-collected dataset [108]. Authentic images were exclusively those that the SREFI algorithm was trained on [108]. Across the three experiments participant accuracy varied little: 49.1% ($N$=176), 49.7% ($N$=174), and 49.7% ($N$=172). These results are ten percentage points lower than those reported in the current study, but one reason for this difference could be the type of authentic images used in that study. As noted, none of the FFHQ images were used, and, as far as we can tell, the authentic images used were qualitatively different to them. That is, rather than resembling the sorts of images found on social media platforms (and found in the FFHQ dataset, and consequently StyleGAN2 images), these images had a uniformity to them. For example, they appear to often consist of a person standing with a neutral expression in front of a plain single-colour background. This is a potential experimental confound that could have confused participants and led to poor performance. Ideally, the images in the authentic stimulus pool would be as similar as possible to the images in the deepfake stimulus pool.

The second related study [104] (and to a lesser degree the third [107]) raises the interesting question of whether the dataset for human deepfake detection accuracy studies should be curated or simply taken as random output from the algorithm. The authors of that study [104] reason that an attacker would be able to curate the dataset they use, since they can see "obvious" rendering artefacts and simply exclude those from their selection. How easy this would be is debatable. Moreover, a crucial difference between the use of deepfakes and shallowfakes is the sheer scalability of attacks that use deepfakes: since a large number can be made in a short amount of time, they can be used for bulk attacks. A curation step is possible, but this would decrease the scalability of attacks. However, our study also shows that for deepfake images with arguably very "obvious" rendering artefacts participants did not have great accuracy even when given the advice interventions. Additionally, since we took output straight from the algorithm and did not curate it, our findings suggest that bulk attacks are surprisingly viable – a finding not evident from prior literature.

In other previous work discussed in the introduction [61], poor performance might be explained by the fact that participants had to respond quickly and had to rate a large number of images which is likely to have resulted in fatigue. In our study, participants took a mean of 78.28 seconds to answer each question (SD = 123.54 seconds), which is much longer than the 5.14 seconds taken by participants in one previous study [61]. Moreover, in our study, each participant was only asked about twenty images (compared with 1,000, 128, and 50 images in the three previous relevant studies [61,104]). As such, the poor performance observed here is unlikely to be due to cognitive overload or haste. Participants were just not very good at the task.

Of further concern is the fact that we found little evidence to suggest that participant's confidence in their judgements correlated with their accuracy. If there had been a positive correlation, then the consequences of participants' low accuracy at deepfake detection would not be so bad, as people would have lower confidence when their accuracy was lower, and so would be less inclined to assert that their incorrect answer was correct. However, the lack of such a correlation as well as the uniformly high confidence given by participants (overall mean = 6.8 on the 0-9 Likert scale) indicates that any deepfake/real label decisions that people make about images of human faces will be likely accompanied by a high degree of confidence regardless of whether the answer is correct or not. There is an extensive literature on overconfidence and its dangers in the psychology literature [129]. Where deepfakes are concerned, human overconfidence is particularly dangerous as there are currently no implemented technical measures or methods by which a person can verify whether an instance is a deepfake or not, and so all a person has to go on is their own intuition – in which they apparently tend to have a great deal of confidence.

Considering variation in the effectiveness of the advice provided, we found that some "tell-tale" signs were used more than others and, more importantly, some were more discriminatory than others. For example, use of the advice regarding the asymmetry of earrings resulted in a rate of about 8 correct deepfake detections to only one false positive. The advice regarding "Strange Clothing Fabric", "Asymmetric Glasses", and "Colour Bleeds From Background" also produced good rates of true-to-false positive judgements (5.17, 4.36, and 2.07, respectively). Considering efficacy, these four items were used by participants who were provided with advice about them for 27.26% of their responses. The reason for their relatively low use is likely to do with the fact that these features may not be present in all images. That this is the case would limit their utility in the real world. In addition, an attacker could remove two of the features (Asymmetric Earrings and Asymmetric Glasses) using an image editor such as Photoshop or GIMP, or simply exclude images that contained these features.

A further concern is that real-world conditions may be less favourable than those employed here. To explain, the image size used in our experiment was set so that pictures would take up the majority of the screen when viewed on a normal laptop (the image size was set to be 720 by 720 pixels so that it would be displayed identically to each participant regardless of screen size). This means that participants would have likely viewed the images in a much larger format than such images would commonly be viewed in real life. For example, profile pictures on social media platforms will generally take up a much smaller portion of a screen; with dating apps, images take up the whole of a smartphone screen, but this size will still be smaller than images were viewed in our experiment. Some platforms also limit image size. For example, Instagram has a profile picture limit of 110 by 110 pixels. The effect of different image sizes upon participant accuracy is an area for future work, but it seems reasonable to suggest that certain details (e.g. the absence of symmetry for earrings) that would affect a participant's decision would be hard or near impossible to see in smaller images. In addition, the rectangular portrait crop that is forced by some smartphone dating apps may obscure some features of deepfakes that might otherwise be observed. Additionally, the efficacy of interventions will be reduced if people do not routinely employ them in real life. However, even if they do, efficacy will be low if they do not work under controlled conditions.

### 4.2. Limitations

As with any experiment, there is a risk that participants assigned to experimental conditions may not have paid attention. In this case, we can likely rule out this threat since those in the advice conditions were found to be more likely to report using the specific advice given when making decisions than those in the control condition.

Additionally, response bias could be an issue, however we address this possibility by computing a Signal Detection Theory measure that takes into account response bias. We calculate $d$ prime for each participant, and an ANOVA conducted over these values indicates no statistically significant differences ($F(3,276) = 2.46$, $p > 0.05$).

This study is not without limitations, however. Chief amongst these is the fact that participants were mostly young (with a mean age of 26.24, $SD = 8.62$), and few reported that English was their first language (9%), with even fewer being UK nationals (5.4%) or US nationals (2.5%). However, the responses provided by participants consistently demonstrated full comprehension of the experimental task and English language fluency. Future work might systematically explore deepfake detection accuracy for different age groups. For example, relative to their younger counterparts, older adults are more often targets of fraud [130], are likely to be less aware of technological developments, and could have worse eyesight and attention to detail. Our expectation, however, would be that the elderly would be less accurate than their younger counterparts.

Another limitation is that the a-priori statistical power analysis was calculated such as to detect a "moderate" (Cohen's $f = 0.25$) effect size across experimental conditions. This means that the experiment would likely not have enough statistical power to detect effects that were "small" in size; if the advice increased participants' accuracy by a "small" effect size, then this would not be detected in this experiment. However, the authors consider practical significance to be important too. That is, using a sample size of 1000 participants, we could perhaps have detected (say) a 2% difference in accuracy between conditions. However, from a policy perspective, achieving such a small effect would be of little value.

The size of the stimulus pool ($N=100$) from which the images were drawn could have been too small for the deepfake stimuli to fully represent the output of the StyleGAN2 method. This size was chosen to allow for a sufficient number of participants to respond to each image for per-image accuracy results to be reliable but this represents a trade-off. The study findings are also limited to the specifics of the interventions tested, these mainly being the number of images shown in the Familiarization intervention (twenty) and the particulars of the Advice provided. Different advice may result in different outcomes which future work might explore.

Finally, the experimental setup employed is unlikely to reflect in-the-wild circumstances, both in that the participants here had full awareness of the presence of AI-generated images in the stimulus pool, and in that participants were explicitly taking part in an experiment. In both cases, we would expect this to mean that participants in the experiment would perform better than those in a real-world scenario, meaning that our estimates may be – if anything – on the optimistic side. Of course, judgements made in the real-world can have real consequences and this too may impact upon their accuracy, the confidence people have in their judgements, and the subsequent choices they make. In the current study, we used an (albeit small) economic incentive to encourage participants to pay full attention to the task. Whether this had the intended effect is hard to know but the amount of time participants took to make decisions (again, a mean of 78 seconds per decision) and the frequency with which they reported items from the Advice intervention (see Figure 3 and Figure 4) suggests that they did so.

### 4.3. Future Work

The limitation of this study to the use of the StyleGAN2 deepfake generation method prompts the need for comparative future work with other deepfake generation methods, of image modality or other modalities. Other versions of the Familiarization or the Advice intervention could be investigated, using a subset of the advice items, different advice, or other types of Advice intervention such as explanatory or demonstrative video segments, or cartoon strips [131]. A different direction should also be investigated: giving participants advice on how to tell that an image is a real image, that it is not a deepfake. This could be fruitful advice since the deepfake generation algorithms find certain things difficult, like brand logos or other text, or objects that partially obscure the face.

An attacker could hand-forge these elements into deepfake images, however. Therefore, future work should also test the impact of basic and quick image editing of the deepfake image stimuli in Adobe Photoshop [85] (or GIMP [86], a cost-free alternative) on participant accuracy. Any successful human interventions or technological solutions should pass this test too: such image-editing is available to an attacker, albeit at the cost of a small amount of time and effort per image.

## 5. Conclusion

The findings of this study suggest that people are not naturally good at detecting deepfakes and that the simple interventions tested do not help. Unfortunately, participants tended to be confident in their ability to differentiate real and deepfake images but their confidence was misplaced. This is a cause for concern and emphasises the threat that the misuse of deepfake images poses.

The StyleGAN2 deepfake generation method whose output this study has assessed has been expanded such that the images can be altered within the StyleGAN2 algorithm. The result is that the same facial identity can be retained while adjustment is made to elements including head pose, baldness, lipstick, background, illumination and face shape [132–134]. These adjustments can also be applied to a picture of an existing person [135]. The fact that this adjustment is made within the StyleGAN2 method (in its "latent space") means that the resultant images will be of the same quality as the regular output of StyleGAN2; in fact, the images can be carefully tweaked to reduce the appearance of visual tell-tale features (e.g. an "Impossible Location" background can be blurred).

These developments expand the potential harms of this technology. For example, a portfolio of facial images can be made with differing expressions, apparent contexts, and even apparent age. This could be put to use to create fraudulent dating app or social media accounts, and the potential to use an existing face as input means that identity theft and a plethora of consequent harms have also become possible. Use of a portfolio of multiple images means that the individual images may no longer carry such a burden of veracity: if people are not aware that several images of a human face could be created with differing expressions, they may be even less likely to consider that what they are looking at is a fake.

This expansion of the StyleGAN2 method could also be combined with deepfakes in the audio modality, which can take a sample input voice and generate samples of that voice's identity saying custom text [136,137]. This could be used to strengthen a fake portfolio of an existing or a non-existent personal identity to enable further harms. This audio deepfake technology is available via an existing company [138] but is also freely available as code on GitHub [139].

The question of having the source code for deepfake generation technologies available for download from GitHub is a debate that should be framed in terms of Responsible Research and Innovation [140]: the potential societal impacts should be considered and taken into account. On the one hand, having the code available on GitHub allows researchers to create and test defence mechanisms for these deepfake technologies, as has been done in this study. On the other, having the code publicly available allows anybody to download it and use it for malicious purposes, for which suitable defences may not yet have been developed. Perhaps a better solution would be to have the authors provide the code on request, if the requester provides a suitable reason and easily-authenticatable credentials (for accountability). In this way, there would at least be an additional barrier to entry for would-be-attackers to overcome.

Perhaps more concerning is the fact that at the time of writing there is a public-facing website that publishes a new deepfake instance (of the StyleGAN2:FFHQ type) every two seconds [116]. A host of similar websites [115,124] additionally publish new deepfake instances from the StyleGAN2 method as trained on datasets with different content (e.g. UK MP headshots [141]). These websites are not hosted or maintained by the authors of the StyleGAN2 publication, but by various private individuals who will have downloaded the StyleGAN2 code from GitHub, trained a StyleGAN2 model on the given dataset, and routed the model's output to the given website. In terms of Responsible Research and Innovation [140], there is a potential argument for this activity: such websites spread awareness of the various capabilities of this deepfake technology, and may in this way be helping to alert the public to the potential threats this technology may cause.

However, such awareness could easily be achieved by hosting a small dataset (e.g. $N$=20) that showcases the output of such deepfake generation methods. An argument against these websites is that their maintained presence may have unintended consequences. There is a difference between the StyleGAN2 code being hosted on GitHub, and the output from a carefully-trained StyleGAN2 model being published continually to a public website. In the first case, several barriers of effort and resource stand between an attacker and a well-crafted deepfake instance: implementing the code requires a degree of programming proficiency and understanding; a dataset of the desired type of instance (e.g. photo of an MP) needs to be sourced, and sifted either automatically or manually, cropped, and aligned; and expensive computing equipment is then needed to train the StyleGAN2 model, all before any deepfake output is created. In the second case, none of these barriers exist, and all that is required for malicious application is the imagination to figure out how to exploit the given deepfake instances. Given all of this, it may not be surprising that several deepfake instances (of the StyleGAN2:FFHQ type) have been found in malicious applications in the wild [121–123]. Open

Science should be embraced [142], but must be accompanied by an appropriate degree of responsibility [143].

**Funding**

This project was funded by the UK EPSRC grant [EP/S022503/1] that supports the Centre for Doctoral Training in Cybersecurity delivered by UCL's Departments of Computer Science, Security and Crime Science, and Science, Technology, Engineering and Public Policy.